# Exploring Application Performance on Emerging Hybrid-Memory Supercomputers

Ivy Bo Peng, Stefano Markidis, Erwin Laure
Department of Computational Science and Technology
KTH Royal Institute of Technology, Sweden

Gokcen Kestor, Roberto Gioiosa
Computational Science and Mathematics Division
Pacific Northwest National Laboratory, USA

***Abstract*—Next-generation supercomputers will feature more hierarchical and heterogeneous memory systems with different memory technologies working side-by-side. A critical question is whether at large scale existing HPC applications and emerging data-analytics workloads will have performance improvement or degradation on these systems. We propose a systematic and fair methodology to identify the trend of application performance on emerging hybrid-memory systems. We model the memory system of next-generation supercomputers as a combination of "fast" and "slow" memories. We then analyze performance and dynamic execution characteristics of a variety of workloads, from traditional scientific applications to emerging data analytics to compare traditional and hybrid-memory systems. Our results show that data analytics applications can clearly benefit from the new system design, especially at large scale. Moreover, hybrid-memory systems do not penalize traditional scientific applications, which may also show performance improvement.**

## I. Introduction

Memory is a central component of every computing system, from small embedded systems to large-scale supercomputers. Among various employed technologies, dynamic random access memory (DRAM) has certainly gained and will continue gaining a major role in storing data. Currently, DRAM represents an effective technology for fast read/write transfer, relatively low cost and fairly large capacity. Supercomputers have employed DRAM as the main memory for decades while other memory technologies, such as processor caches, have been used to support access to DRAM.

The balance between DRAM and other memory technologies is about to shift, mainly due to four factors. First, the scaling factor of DRAM technology is progressively reducing to a point that a hard stop is expected in the near future. Among various issues, the increase of leakage power and the reliability of the manufacturing process have become major problems for scaling DRAM technology. Second, emerging irregular HPC applications and data analytics workloads require an ever-increasing amount of memory to store data. The density and static power consumption of DRAM make it an unviable solution for these new workloads. Memory technologies with higher density and lower power consumption are required. Third, the massive number of cores and hardware threads in modern processor chips need to be sustained by a higher bandwidth than the currently DRAM technology. Finally, previous studies have pointed out that persistence will greatly help reduce the energy cost due to data movement between computational applications and data analytics and visualization tools [1].

All these factors have indicated that the days of DRAM being the sole dominant technology employed on supercomputers are near the end. However, there is currently no single new technology that can completely replace DRAM by providing the necessary bandwidth, density, storage size, and low energy profile at the same time. New volatile technologies, such as 3D-stacked DRAM, employ re-designed circuit and interface and can achieve faster and efficient data transfers [2], [3]. Their high cost limits their use to small quantities insufficient for the large data structures in data analytics workloads. Emerging non-volatile memory (NVRAM) technologies, such as Phase-Change Memory, Spin-Transfer Torque RAM and Resistive Random-Access Memory, have shown potential of delivering persistent, high-density and low-power memory at a relatively low cost [4], [5], [6]. Nevertheless, the current NVRAM technologies still have limitations of high access latency, asymmetric read/write bandwidth, and low write endurance compared to DRAM.

Next-generation supercomputers are likely to feature a deeper and heterogeneous memory hierarchy, where memory technologies with different characteristics work next to each other, mitigating each other's drawbacks. For example, the Intel Xeon Phi Knight Landing processor features 8-16 GB 3D-stacked Multi-Channel DRAM (MCDRAM) in addition to 384 GB of "far" conventional DDR4 RAM to support memory-intensive applications [7]. Similarly, the Summit supercomputer to be delivered in 2017 will feature 500 GB DRAM and 800 GB slower NVRAM [8], [9]. This new trend is in net contrast with traditional supercomputers that employ DRAM technology for main memory storage, besides caches and accelerator on-board memory.

A critical question for hardware and software designers is whether this new computing paradigm, where compute nodes feature multiple memory technologies side-by-side (hybrid-memory systems), performs better than the traditional scheme using DRAM as the main memory. In this paper, we try to quantitatively answer this question with a systematic and fair methodology at scale using both traditional and emerging workloads. Without loss of generality, we assume that a supercomputer can either be equipped with a "fast" memory or with a combination of a "fast" and a "slow" memories.

Systems with multi-level memories can be derived from this scheme considering point-to-point transfers between any two kinds of memory. We also assume that both fast and slow memories are byte-addressable. Block-addressable memories connected to I/O buses and accessed through file system interfaces pose entirely different challenges. The generally large overhead of those block-addressable memories makes them infeasible for a direct replacement for main memory. We then consider two system designs: a traditional Uniform-Memory System ($UMS$) equipped with $M$ GB of fast memory and an alternative Hybrid-Memory System ($HMS$) equipped with $M$ GB of fast memory placed next to $N$ GB of slow memory. To provide a fair comparison, we configure the two systems with a number of compute nodes such that the total amount of memory in $UMS$ and $HMS$ is the same. This model can be used to represent systems like the Intel Knight Landing with fast MCDRAM memory and slow DRAM memory, as well as the nodes on the coming Summit and Aurora supercomputers equipped with fast DRAM and slow NVRAM [7], [8], [9].

We performed our experiments on a cluster of 128 compute nodes equipped with two 16-core AMD Interlagos processors and interconnected through Mellanox ConnectX-2 InfiniBand high-speed network. Each node features 64 GB DRAM divided into four NUMA domains. We use the NUMA domains to emulate the fast and slow memories in the $UMS$ and $HMS$ system designs. We tested a variety of workloads, from real-life HPC applications like GTC [10] and Lammps [11], to traditional scientific benchmarks, such as conjugate gradient solvers, LU matrix decomposition, and discrete Fourier Transform kernels, to emerging graph benchmarks, such as Graph500 [12] and GUPS. We also extract key applications characteristics and correlate them to the performance improvement or degradation observed on $HMS$ system. To the best of our knowledge, this is the first study showing the impacts at scale of compute nodes with multiple memory technologies on the traditional scientific and emerging data analytics applications. Our experiments show that irregular applications benefit the most from the $HMS$ design, with performance improvements up to 2.74x. Regular, CPU intensive applications with good cache locality only marginally take advantage of the larger memory.

This paper makes the following contributions:

- We propose an effective and low-overhead methodology to emulate hybrid-memory systems at scale.
- We compare traditional and hybrid-memory systems and perform a quantitative analysis using both classic HPC and emerging data analytics applications. To the best of our knowledge, this work is the first to provide such in-dept analysis.
- We extract and analyze the key factors that indicate whether an application benefits from future hybrid-memory system designs and to which extent.

This paper is organized as follows: Section II examine previous related works. Section III describes the methodology. We describe the test system and the applications in Section IV. Section V presents the our results. We discuss the findings and limitations of the current work in Section VI. Finally, Section VII concludes this work.

## II. Background and Related Work

New memory technologies and their use in computing systems have been studied for many years. We classify the related works in the following three categories.

**Simulators:** The impact of novel memory technologies and memory controllers is generally studied with cycle-accurate simulators [4], [13]. These approaches are very accurate but generally require long execution times. Thus, they are practical only for small-scale systems, such as single threaded executions or single compute node with up to tens of cores. Also, they typically require small micro-benchmarks or partial execution of applications. In contrast, our work analyzes the impact of hybrid-memory system on a large scale cluster consisting of 128 compute nodes equipped with a total of 8 TB of memory. Moreover, our evaluation is based on complete, real-life parallel applications and kernels using problem sets stretching the limit of the underlying system.

**Emulators:** To increase simulation speed and perform larger scale studies, researchers have also used hardware and software emulators. Hardware emulators are implemented on accelerators, such as FPGA [14]. This type of emulators can provide accuracy similar to simulators but at a much faster speed. However, these approaches are limited by the size of the FPGA and are still slow compared to native execution. As a result, only small memories and applications can be emulated with this approach. Software emulators are generally based on latency injection to emulate "slow memories". These emulators can be implemented in the operating system (OS) device driver or in a user-level library [15], [16]. The main advantage of software-emulated solutions is that applications run natively on the compute node. The main limitation, however, is that the introduced runtime overhead might be larger than the memory access time of some memory technologies, such as MCDRAM. For this reason, these solutions are particularly effective to emulate NVRAM memories attached to the I/O bus [15]. Although we use a form of low-overhead emulation, we target large scale systems with byte-addressable memory technologies. Thus, neither hardware nor current software emulators can be used for our analysis.

**Hybrid Memory Systems:** Many studies on hybrid memory systems have focused on simultaneously employing DRAM and NVRAM technologies in a single compute node. One main objective of these studies is to mitigate the limitations of NVRAM by using DRAM strategically. There are two main hybrid memory designs: a flat memory space with DRAM and NVRAM placed side-by-side or a hierarchical memory space using DRAM as cache for NVRAM [4], [13], [17]. Researchers have proposed hardware extension, OS support, and algorithm-directed data placement to achieve optimal data movement between DRAM and NVRAM [4], [16], [17], [18], [19]. Our objective in this work is not to design a system to transfer data between DRAM and (I/O) NVRAM, but rather

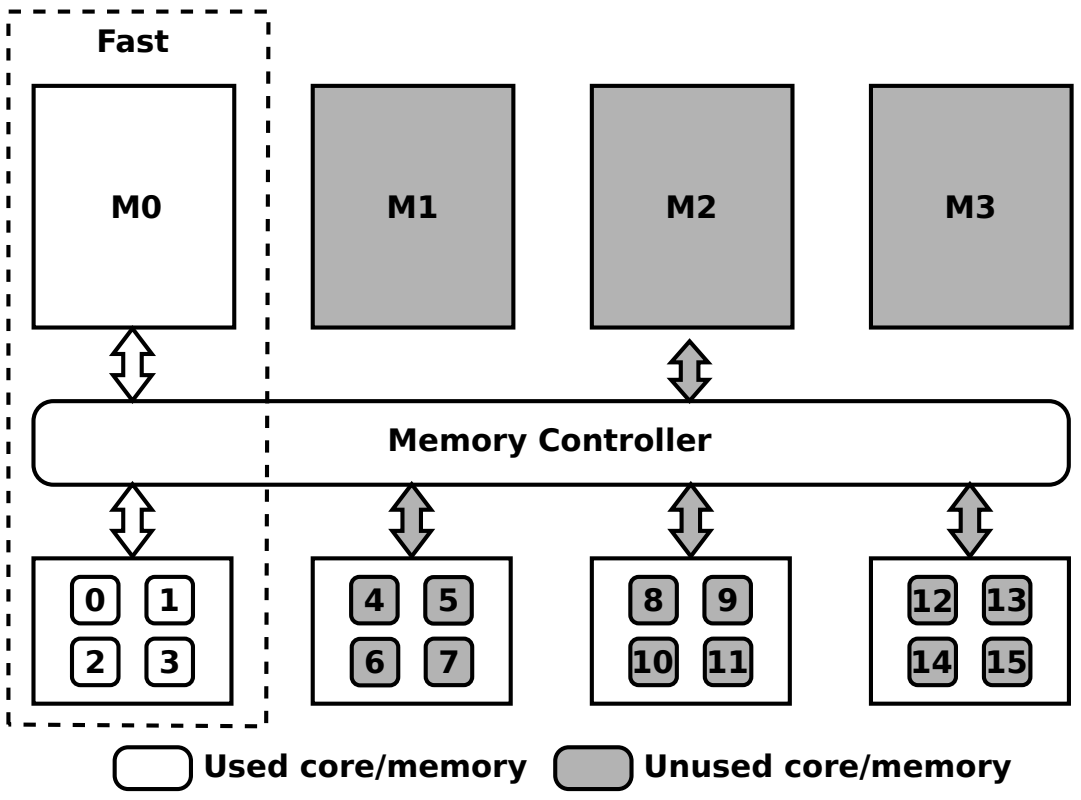

(a) Uniform Memory System ($UMS$)

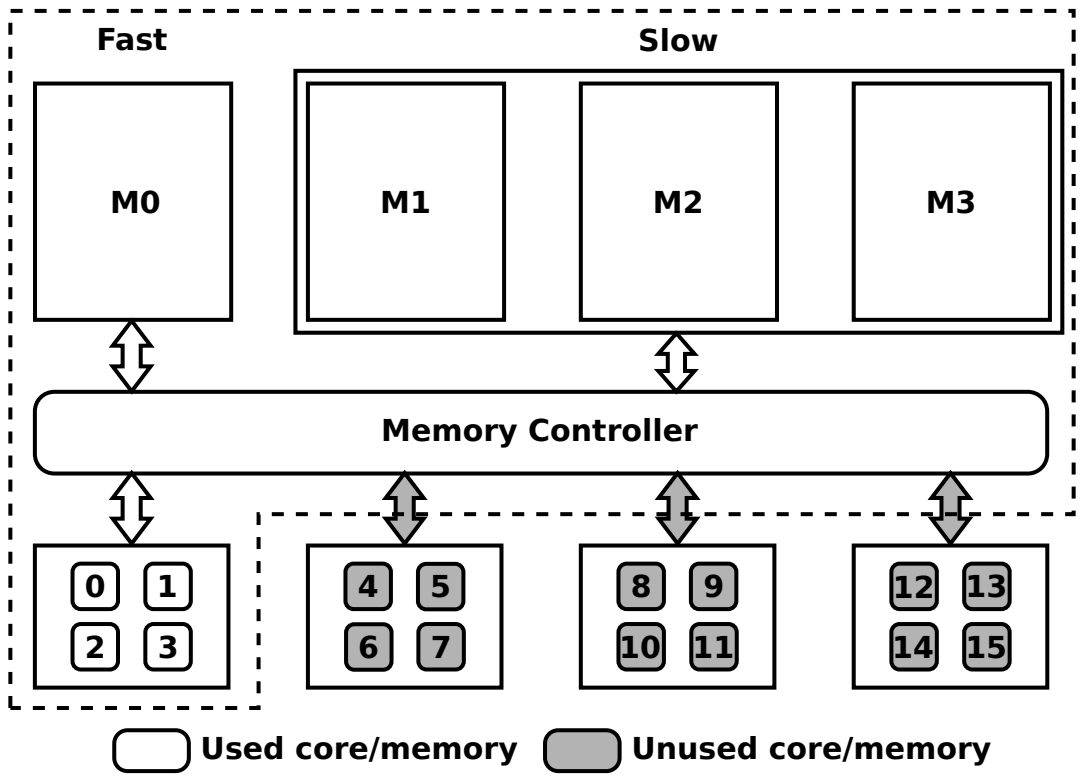

(b) Hybrid Memory System ($HMS$)

Fig. 1. $UMS$ and $HMS$ emulation. Dashed boxes represent the system components in use; shaded components are not used in the experiments.

to analyze the effects of using multiple memory technologies side-by-side at large scale. We also provide an analysis of the application characteristics and identify the key factors for an application to benefit from hybrid-memory systems.

## III. Methodology

Our goal in this work is to analyze the performance of real-life, distributed-memory applications running on supercomputers equipped with multiple memory technologies placed side-by-side. In particular, we are interested in byte-addressable memory technologies that can be accessed with low latency and through the common load/store interface. We refer the reader to previous works on systems that uses DRAM as cache for NVRAM attached to the I/O bus [13], [17].

As discussed in the previous Section, simulation environments are not suitable for experiments at scale nor for large applications. Software emulation environments are suitable for memory technologies attached to the I/O bus, but introduce runtime overhead orders of magnitude larger than the memory access time of byte-addressable memories. Thus, we resort to an emulation strategy that leverages the intrinsic heterogeneity of Non-Uniform Memory Access (NUMA) systems. In NUMA systems, memory is divided into several domains and each core can access its *local* memory with low latency, or the *remote* memory with a larger latency. Figure 1 shows a typical NUMA system design: memory is divided in four NUMA domains and compute cores are divided into four groups of four cores each. Cores can access a local memory and three remote memories. For example, cores 4-7 have fast access to the local memory $M1$ but pay a higher latency in accessing the remote memories $M0$, $M2$ and $M3$.

We use the intrinsic memory latency heterogeneity of NUMA systems to emulate two different system designs: a traditional Uniform-Memory System ($UMS$), in which CPU cores can access all memory with the same latency, and an alternative Hybrid-Memory System ($HMS$) design, in which a "fast" memory is placed next to a "slow" memory. We assume that the memory in $UMS$ can be accessed with the same latency of the "fast" memory in $HMS$. Figure 1a shows how we emulate $UMS$ systems (shaded components are not used during the experiments for $UMS$). We only use a set of cores (0-3) and the corresponding local memory ($M0$). We execute our tests properly binding applications threads to the correct CPU cores and force the allocation of data structures in the local memory through `numactl`. In this configuration, cores 4-15 and memories $M1$-$M3$ are not used, thus the system behaves as a uniform memory system where all cores access all memory with the same latency.

TABLE I
NUMA Distance

| Node | 0 | 1 | 2 | 3 |
|---|---|---|---|---|
| 0 | 10 | 16 | 16 | 16 |
| 1 | 16 | 10 | 16 | 16 |
| 2 | 16 | 16 | 10 | 16 |
| 3 | 16 | 16 | 16 | 10 |

To emulate $HMS$, we perform a different configuration, as shown in Figure 1b. We still use the same set of cores (0-3) but allow cores to access all available memories $M1$-$M3$. In this configuration, $M0$ acts as "fast" memory, while $M1$-$M3$ act as "slow" memories. Note that, in Figure 1b, the memory controller also accesses memories $M1$-$M3$ (white arrow) and that the remote memories are seen as a single, slow memory by cores 0-3. Table I shows the distances between any two NUMA domain, as reported by `numactl --hardware`. In the Table, a value of 10 represents the distance to the local memory and larger values represent larger distances. `numactl --hardware` reveals that cores 0-3 can access $M1$-$M3$ with latency that is 1.6x larger than the latency of $M0$. To ensure that all memories are utilized during the execution of our experiments, we use `numactl --interleave` to force interleaved allocation of the application data structures on $M0$-$M3$. This configuration represents a system consisting of 25% of "fast" memory and 75% of "slow" memory. This ratio is in between systems like Intel Knight Landing (up to $4/96$) [7] and the CORAL systems ($\approx 40/60$) [8], [9].

Obviously, the two system designs have different characteristics, i.e., $HMS$ compute nodes are equipped with four times more memory than $UMS$ compute nodes. To perform a fair comparison and given the ever-increasing demand for system with larger memory, we establish the *same memory principle*, i.e., both system designs should provide the same total amount of memory. Essentially, we are comparing $UMS$ and $HMS$ from the prospective of a user who is interested in running applications that require a certain amount of total memory. To achieve this objective the $UMS$ consists of four times more compute nodes than the $HMS$.

This choice, however, brings another possible inequity: since $UMS$ has four times more compute nodes and assuming that all cores in a NUMA domain are employed, the $UMS$ system will have four times more computing power than the $HMS$ system. We could solve this problem by using only $1/4$ of the compute cores in one NUMA domain for the $UMS$ systems, however, we need to consider that cores in modern processor chips share multiple levels of hardware resources, e.g., all the cores in one NUMA domain share the last level cache (LLC). Using only $1/4$ of the compute cores will increase the per-core LLC in the $UMS$ system (4x) with respect to the $HMS$ system. This could result in considerable performance variation, especially for applications that are capable to exploit cache locality. As a matter of fact, this choice would introduce an heterogeneity between the processors used in the $UMS$ system (one core, full LLC) and those used in the $HMS$ system (four cores, 1/4 of the LLC per core). To solve this issue, we decided to use all cores in a NUMA domain for both the $UMS$ and the $HMS$ designs. However, to abstract from the fact that $UMS$ systems have four times more cores, we compare the two designs in terms of performance per core. This provides a fair comparison, as each core has access to the same amount of LLC, the total number of cores does not affect the evaluation, and the only key differential factor is the memory distribution across the compute node, which is the objective of this study.

## IV. Experimental Environment

### A. Experimental Setup

We performed our evaluation on a cluster of 128 compute nodes interconnected through an Mellanox ConnectX-2 Infiniband network fabric. Each compute node is equipped with two AMD Interlagos [20] 16-core processor chips running at 2.1 GHz and a total of 64 GB RAM. Each processor chip is divided in two modules that comprise eight cores, a shared 8 MB last-level cache, and a memory controller. Thus each compute nodes consists of a total of 32 cores and four NUMA domains (16GB each). When emulating the $UMS$ system, we only use the first NUMA domain (8 cores, 16 GB). Conversely, when emulating the $HMS$ system, we also only use cores in the first NUMA domain (8 cores) but use all four NUMA domains (64 GB). As explained earlier, using fewer than eight cores changes the balance between cores and caches.

All applications are compiled with `gcc` version 4.6.2 and linked against `OpenMPI` version 1.8.4. The cluster runs a standard `Linux` 4.1.0. Hardware performance counters are extracted with the `Linux` 4.1.0 `perf` tool. We perform five repetitions of each experiments and report the average results given less than 2% standard variation.

TABLE II
SUMMARY OF BENCHMARKED APPLICATIONS

| Application Name | Input Parameter | Memory Footprint | Scaling Type |
|---|---|---|---|
| Graph500 | $2^{31}$ vertices graph | 640 GB | Weak |
| GUPS | $2^{37}$ length table | 1100 GB | Weak |
| Lammps | $4.2 \times 10^9$ atoms | 750 GB | Weak |
| HPCCG | $1024 \times 1024 \times 512$ grid | 400 GB | Weak |
| GTC | $1.3 \times 10^{10}$ particles | 1050 GB | Weak |
| FT | $2048 \times 1024 \times 1024$ grid | 120 GB | Strong |
| LU | $408 \times 408 \times 408$ grid | 13 GB | Strong |

### B. Application Workloads

Traditionally, scientific applications have used the vast majority of cycles on HPC platforms. Recently, data analytics workloads have started consuming more cycles as only large HPC systems provide enough memory to store the data sets. To provide an exhaustive analysis, we analyzed both traditional HPC applications and data-analytics workloads. We selected two NAS parallel benchmarks, Fast Fourier Transform (FT) and Lower-Upper symmetric Gauss-Seidel (LU), one parallel preconditioned conjugate gradient solver (HPCCG), two data-intensive applications, Graph500 and the HPCCG random access benchmark (GUPS), and two real-life applications, a modular dynamic application (Lammps) and a particle-in-cell code (GTC). The input parameters, the total memory footprint, and the type of scaling tests of each applications are summarized in Table II. We chose a mix of weak and strong scaling tests. For weak scaling applications, the input set is chosen to use all available memory (Table II show the memory footprint at the largest scale). For strong scaling applications, the input set is chosen to fit in the smallest configurations (one node for $HMS$, four nodes for $UMS$).

FT benchmark calculates forward and inverse DFTs. Each process sends different data to each other process through a all-to-all communication to transpose the distributed matrix. 1D DFT is then performed on one dimension with contiguous memory access. LU is a pseudo computational fluid dynamic application solving lower and upper triangle systems. It is computational intensive and presents low communication to computation ratio. HPCCG calculates the eigenvalues of an unstructured sparse matrix. It presents random memory accesses and requires both sparse and collective communication.

The Graph500 [12] benchmark implements a level-synchronous breadth-first search (BFS) that is fundamental for many data-analytics applications. The application uses a Kronecker generator to generate very large graphs. The communication is driven by the structure of the graph, exhibiting an irregular pattern. GUPS is a synthetic benchmark that measures the Giga-updates-per-second by reading and updating random addresses in a global table distributed over all processes. Communication is required to update the part of table located on a remote process. The memory access and communication pattern are irregular and random.

Lammps [11] is a classical modular dynamic code that calculates the forces among atoms. Given a large number of atoms, Lammps is computation intensive. The communication pattern is regular and mostly sparse, limited to neighbours processes. Molecular dynamics is one of the most important class of simulations running on the current supercomputers. GTC [10] is a magnetic fusion application that represents another important class of applications. GTC employs a particle-in-cell technique that simulates ions and electrons with computational particles and solves the gyro-average Vlasov equation on a computational grid. It consists of a computation intensive part that updates the position and velocity of each particle and an iterative solver for calculating the field values.

## V. Evaluation

In this section, we present a quantitative comparison of $UMS$ and $HMS$ designs using a set of scientific and data-analytics applications. We also correlate the observed performance improvements to applications key characteristics.

### A. Application Performance

We provide a quantitative comparison between the $HMS$ and $UMS$ system in terms of per-process performance of the seven applications using the methodology described in Section III. Figure 2 shows per-process performance measured with application-specific metrics for both $HMS$ and $UMS$ (red and blue bar, respectively) and the $HMS$ improvement relative to $UMS$ (in black line). For GUPS, we report the number of updates/second/process; for Graph500 the number of traversed edges/second/process (TEPS/process); for Lammps tau/day/process; for GTC the number of processed particles/second/process; finally for FT, LU, and HPCCG, we report the number of floating-point operations/second/process (FLOPS/process). The x-axis shows the number of processes used in the $HMS$ and $UMS$ systems. For example, 64:256 indicates that the experiments on the $HMS$ system used 64 processes (8 compute nodes, 64GB of memory each), while the equivalent $UMS$ system used 256 processes. (32 compute nodes, 16 GB of memory each). We remark that the total memory used is the same on the two systems.

The data analytics applications (GUPS and Graph500 in Figure 2a and 2b) present the largest improvements. The $HMS$ system achieves up to 2.74x and 1.31x improvement over the $UMS$ system for GUPS and Graph500. The results of these two applications show that irregular memory access and communication patterns can be mitigated when using compute nodes with larger memory. For Graph500, the next vertex to be accesses is only known after the edges of the current vertex have been examined. The probability of accessing a remote vertex depends on the graph topology and partitioning, thus there is a higher probability of accessing a remote vertex in the $UMS$ system because of the smaller memory of each compute node. Moreover, the graph structure of real-world networks is often a scale-free network following a power law in their degree distribution. In such scenario, there are only a few "hubs" with high degree in a graph while other vertices only have a few edges. If a "hub" is allocated to a specific node, it is likely that most other processes try to communicate with this node. For the set up of our experiments on 32 $HMS$ compute nodes, only 31 nodes communicate to one node while on 128 $UMS$ compute nodes, 127 nodes communicate to one node. In such scenario, the $HMS$ system has a lower probability of network congestion compared to $UMS$ system. Obviously, 75% of the $HMS$ node memory is slow, thus there is a latency penalty when accessing vertices stored in the slow memory, but this penalty is much smaller than the cost of accessing vertices stored on remote compute nodes through the network. We observe in Figure 2b that the $UMS$ system performs better at small scale (up to 32:128 processes) but than there is a sudden performance drop when using 64:256 processes. On the other hand, $HMS$ performance reduction is less pronounced, hence the overall improvement increases with the number of processes. The black line in Figure 2b shows a clear increasing trend, thus we expect the relative improvements of $HMS$ over $UMS$ to continue increasing at larger scale. The access pattern irregularity is even more evident with GUPS, which performs random memory accesses. When comparing the two systems at the largest scale, i.e., 256 processes running on the $HMS$ system (32 compute nodes) against 1024 processes running on the $UMS$ system (128 compute nodes), the probability that the next memory location to be accessed is on the local node is 3.125% (1/32) for the $HMS$ system and 0.781% (1/128) for the $UMS$ system, assuming uniform random updates. To the contrary of Graph550, we observe that GUPS improvement is not linear with the number of processes but rather shows a random behavior. This is the consequence of the randomness nature of the benchmark. GUPS generates a random table at each execution and there is no dependence between the memory address to be accessed. However, we also observe that there is a considerable performance improvement in all cases, between 1.31x and 2.74x.

The three scientific kernels (FT, LU and HPCCG, shown in Figures 2c, 2d, and 2e) present similar trend in the relative improvements indicated in the black lines. In all cases, $HMS$ improvements over $UMS$ increase as the number of processes increases. At the largest scale, the $HMS$ system achieves similar improvements for FT and LU (1.13x and 1.15x, respectively). More in details, we observe that FT performance (Figure 2c) slowly decreases with the number of processes but that the degradation is more profound on the $UMS$ system. Although we do not observe a monotonically increasing improvement, the number of FLOPS/process shows a fairly clear path for both the $HMS$ and the $UMS$ systems. LU also shows a generally increasing trend in the performance improvement but with a slight performance dip when using 4 $HMS$ compute nodes (32 processes). We notice that LU benefits from the $HMS$ system design much later than FT, only beginning at 8 nodes (64 processes). LU is a compute-intensive and exhibits regular patterns and good locality. At a small scale, these characteristics outweight off-node communication and the performance penalty of accessing slow memory is significant. For HPCCG, the $HMS$ does not outperform

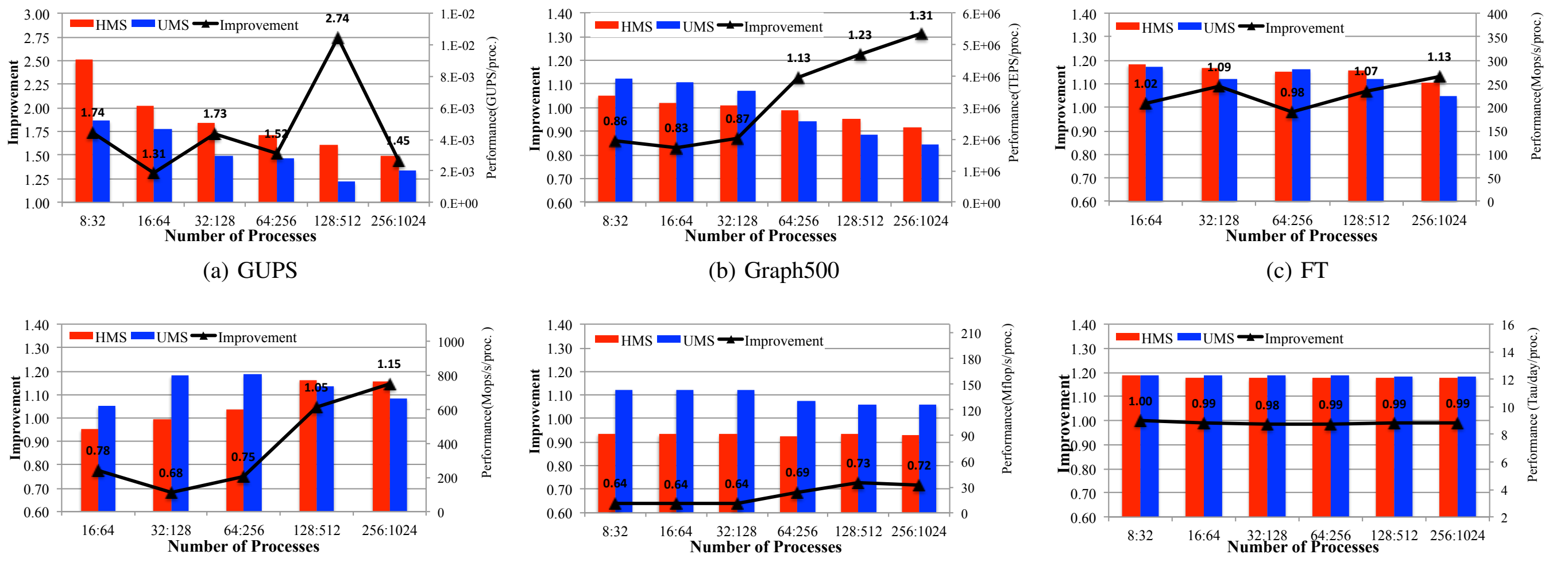

(a) GUPS (b) Graph500 (c) FT

(d) LU (e) HPCCG (f) Lammps

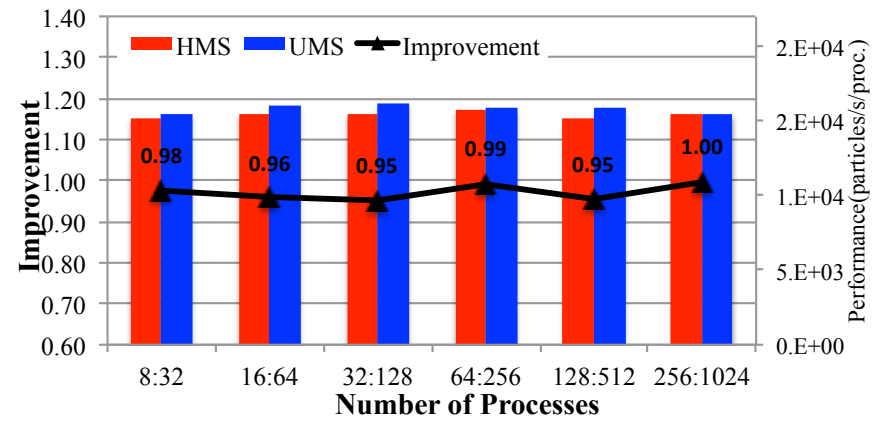

(g) GTC

| HW Counters | GUPS | Graph500 | FT | LU | HPCCG | Lammps | GTC |
|---|---|---|---|---|---|---|---|
| Frontend stalls (%) | 26.58 | 19.65 | 18.37 | 1.92 | 3.13 | 3.48 | 1.66 |
| Mem. prefetch/s | $2.82{\cdot}10^6$ | $2.40{\cdot}10^6$ | $3.67{\cdot}10^6$ | $1.47{\cdot}10^7$ | $1.13{\cdot}10^7$ | $3.50{\cdot}10^6$ | $5.19{\cdot}10^6$ |
| LLCmiss/KLoads | 57.9 | 9.1 | 18.1 | 11.6 | 41.5 | 2.7 | 1.8 |
| Instr/LLCmiss | 38.1 | 197.6 | 128.2 | 159.2 | 52.7 | 681.7 | 1294.3 |

(h) Hardware performance counters

Fig. 2. The $x$ axis indicates the number of processes on $HMS$ and $UMS$ system. Application performance per process is reported in red and blue bars for $HMS$ and $UMS$. The black line indicates $HMS$ improvement over $UMS$. Table 2h reports hardware counters collected with Linux `perf`.

$UMS$ at the largest tested scale. HPCCG is an interesting case, as it presents both irregular data structures (sparse matrix) and near-neighbor communication, which tends to be relatively regular. Compared to the FT and LU, HPCCG is less compute-intensive and has poor data locality due to the unstructured sparse matrix. Given the increasing trend in the performance improvement, we believe that $HMS$ will eventually outperform $UMS$ at larger scales.

Lammps and GTC are two real-life applications representing important scientific codes. These applications have been heavily optimized for current systems ($UMS$). Their performance results show that these two classes of applications perform almost equally on $HMS$ and $UMS$ systems. Both applications demonstrate good scalability on both systems with $HMS$ showing minor performance degradation: less than 2% for Lammps and less than 5% for GTC. The experiment results show that the new computing paradigm will continue supporting the performance of these important scientific simulations.

### B. Application Analysis

Overall, our results in Figure 2 show that there is generally an advantage in using a system with fewer compute nodes but larger memory per node ($HMS$) over a system with more compute nodes equipped with less memory ($UMS$), albeit some part of the memory in $HMS$ is slow. The reasons for these results are multiple and depend on the intrinsic characteristics of each application. In this Section we identify and isolate these application characteristics by analyzing the execution properties of each benchmark. Our objective is to classify applications in terms of irregularity, memory locality, and CPU intensity. This classification and the collected metrics can be used to foresee the impact of using hybrid-memory solutions on a general application.

TABLE III
APPLICATION CLASSIFICATION

| Application | Regularity | Locality | CPU Intensity | Max Improv. |
|---|---|---|---|---|
| GUPS | Irr. | Poor | Low | 2.74x |
| Graph500 | Irr. | Fair | Med | 1.31x |
| FT | Irr. | Fair | Med | 1.13x |
| LU | Reg. | Fair | Med | 1.15x |
| Lammps | Reg. | Good | High | 0.99x |
| GTC | Reg. | Good | High | 0.99x |
| HPCCG | Reg. | Poor | Low | 0.72x |

We analyze the application performance counters and correlate the results with the performance and improvements measured in Figure 2. Figure 2h reports the values of the performance counters collected for each applications. We present normalized results, as each application runs for a different amount of time and performs different operations (e.g., LU, HPCCG, and GTC perform floating-point operations, while GUPS and Graph500 perform integer operations). We only report the counters significant for the discussion in this paper.

Irregularity depends on the data structure, memory access and communication patterns. There is no direct measurement for irregularity, however, we believe that a high percentage of front-end stall cycles and a reduced memory prefetching activity are good indications of irregularity. Front-end stall cycles indicate that the core is stalled because of a data dependency, i.e., one or more of the operands of the scheduled instruction are not ready. Modern out-of-order processors with data pre-fetchers have greatly reduced the amount of time wasted in waiting to resolve data dependencies. Nevertheless, not all dependencies can be eliminated or hidden. For example, if an instruction needs an operand not found in the cache, the core cannot execute the instruction until the operand is moved from memory to cache. Applications with good data locality (useful data resides in the cache) and regular access patterns (patterns recognizable by the data pre-fetcher) are able to hide this latency and resolve the data dependencies before the instruction is scheduled, avoiding processor front-end stalls. Conversely, applications with irregular access patterns that cannot be identified by the pre-fetcher and/or with poor data locality, often stall at the processor front-end. Processor cores can also stall because the scheduled instruction needs an operand computed by a preceding long-latency instruction, such as a floating-point division. Thus, we also consider the data pre-fetcher activity as an indication of the application regularity. However, we remark that applications with a low number of memory loads (more precisely, a low number of cache misses) may not provide enough information to kick the pre-fetcher, hence data pre-fetching must be correlated with the number of loads executed.

Data locality is derived by the number of cache misses per load instruction. We are particularly interested in the data transfers between memory and processor, thus we analyze the last-level cache (LLC) misses per load operation. Load instructions that miss in one of the lower-level caches but hit in one of the higher-level caches do not trigger data transfers, thus are omitted from this analysis. Applications with a high LLCmiss/load have poor locality, as a large number of load operations miss in all processor cache and need to be fetched from memory. Note that an application may have relatively poor locality (data re-used in the cache line is low) but a regular memory access pattern. In this case, however, the data pre-fetcher is generally capable of reducing the number of the LLC miss.

Finally, CPU intensity is computed as the number of instructions (Instr.) per memory transfer (LLC miss). Applications that perform many operations per memory transfer are considered CPU intense and less sensitive to system memory hierarchy and network topology. Conversely, applications with a low number of instruction per memory transfer are classified as memory intensive. These applications are particularly sensitive to memory latency and remote data transfer.

Using the performance counters reported in Figure 2h, we classify applications from the most sensitive to memory hierarchy to the least sensitive and correlate this information to the maximum improvement observed in Figure 2. Table III shows our results. GUPS, Graph500, and FT show a large percentage of front-end stalled cycles and a relatively low pre-fetching activity. These applications are classified as irregular. Note that, although Lammps also shows low pre-fetching activity, its LLC miss to load ratio is also very low. With few LLC misses, the pre-fetcher does not have enough information to attempt autonomous data transfer. A similar reasoning holds for GTC. GUPS and HPCCG are the benchmarks with the lowest data locality (highest values of LLC miss/loads). This is expected given the sparse matrix used by HPCCG and the random memory updates in GUPS. Graph500, FT and LU show a fair data locality, better than GUPS and HPCCG but worse than GTC and Lammps. Finally, GTC and Lammps are the most CPU intensive (highest values of Instr/LLCmiss). At the other side of the specturm, GUPS and HPCCG are considered memory intensive (lowest values of Instr/LLCmiss), while LU, FT, and Graph500 are in the middle.

We observe that irregularity is a key point: all irregular applications perform better on $HMS$ than on $UMS$. Second in order of importance is the application's locality. Applications with poor locality need to access memory more often, hence are more sensitive to the memory hierarchy structure and data distance. Finally, CPU intense applications are less impacted by the data transfers because they spend most of their time performing computation within the processor. The shaded area in Table 2h highlights the applications that benefit the most from hybrid-memory systems and their key characteristics in order of importance. Irregular applications with poor locality and low/medium CPU intensity (GUPS, Graph500) are at the top, while regular, CPU intensive applications with good locality are at the bottom (Lammps, GTC). FT and LU seat in between these two categories. However, we did observe in Figure 2 that FT (irregular) performance on $HMS$ are higher than $UMS$ in almost all the cases, while LU (regular) performance are better on $HMS$ only after 128:512. Performance counters confirm the HPCCG interesting behavior observed in Figure 2e. This application presents irregular memory access pattern (poor locality, low CPU intensity) and a regular communication pattern. We classify the application as regular based on the percentage of front-end stalls but we also observe a high pre-fetching activity. However, as indicated by the high LLCmiss/loads and the low instructions/memory transfer, pre-fetched data does not seem useful.

## VI. Discussion and Limitations

In this work, we leverage the heterogeneous memory latencies of NUMA systems to emulate a systems with different memory technologies are placed side-by-side. This approach allows us to efficiently emulate the higher latency of slow memory at large scale but come with some limitations. First, we cannot change the latency of the fast and slow memories. Even though some systems allow the users to change the latency of accessing NUMA domains, we do not have enough of such compute nodes to make a reasonably large-scale cluster. Second, our fast/slow memory ratio is fixed to 25/75. This value is actually in between systems like Intel Knight Landing and the ones from the CORAL procurement. We could change the ratio to 50/50 (two NUMA domains) or 33/66 (three NUMA domains), but we do expect that the amount of slow memory will be predominant due to its higher density. Finally, read and write latency are equal. This is not representative for some NVRAM technologies that present asymmetric latencies (typically write accesses take longer). Previous studies show that read accesses are predominant in scientific and data analytics workloads, thus our results are valid to the extent that the application with high read/write ratio [16]. Moreover, modern processor caches considerably filter the number of writes that propagate to memory because they typically employ write-back policies.

We do not analyze the effects of the network congestion due to space constraints. It is well-known that irregular network communication can lead to high congestion in the network switches and that barrier/collective operations may limit scalability. Both effects lead to the conclusion that reducing the number of compute nodes will reduce the network congestion because of the fewer messages in flight (see Section V-A). Moreover, using fewer compute nodes could reduce the latency of barrier and collective operations as the communication cost for such operations increases with the number of nodes.

In summary, we believe that our results are valid and representative for hybrid-memory systems featuring different byte-addressable memory technologies side-by-side, such as the Intel Knight Landing processor, as well as systems equipped with byte-addressable NVRAM. We also believe that our application analysis (Section V-B) is general and applicable to other system designs.

## VII. Conclusions

The ever-increasing demand for memory by scientific applications and data-analytics workloads is driving a design shift to the next-generation supercomputers. Future systems will feature hybrid memory technologies organized side-by-side. In this paper, we analyzed the impact of the emerging hybrid-memory system design at large scale on a diverse set of applications from traditional HPC applications (Lammps, GTC, FT, LU and HPCCG) to emerging data analytics workloads (Graph500 and GUPS). Our results show that irregular applications with poor locality are the ones that will benefit the most from larger memories within compute nodes. Moreover, we show that traditional, computing intensive applications may also benefit from this new design and will not be penalized.

To the best of our knowledge, this is the first study that quantitatively compare different system designs at large scale for both traditional and data analytics workloads. Nevertheless, we will extend our work to cover system designs and technologies currently not emulated.

## Acknowledgment

This work was funded by the European Commission through the EPiGRAM (grant agreement no. 610598) and SAGE projects (grant agreement no. 671500). This work was supported by the DOE Office of Science, Advanced Scientific Computing Research, under the Argo project (award number 66150) and the CENATE project (award number 64386).